\documentclass[pra,twocolumn,preprintnumbers,
amsmath,amssymb,superscriptaddress]{revtex4}

\usepackage{graphicx}

\newcommand{\ket}[1]{| #1 \rangle}
\newcommand{\bra}[1]{\langle #1 |}

\begin{document}

\title{Optical Quantum Computation with Perpetually Coupled Spins}
\author{Simon~C.~Benjamin}
\email{s.benjamin@qubit.org}
\affiliation{Department of Materials, Oxford University, Oxford OX1 3PH, United Kingdom}
\affiliation{Centre for Quantum Computation, Department of Physics, Oxford University OX1 3PU, United Kingdom}
\author{Brendon~W.~Lovett}
\affiliation{Department of Materials, Oxford University, Oxford OX1 3PH, United Kingdom}
\author{John~H.~Reina}
\altaffiliation[On leave of absence from ]{Centro Internacional de F\'isica, A.A. 4948, Bogot\'a, Colombia}
\affiliation{Department of Materials, Oxford University, Oxford OX1 3PH, United Kingdom}
\affiliation{Centre for Quantum Computation, Department of Physics, Oxford University OX1 3PU, United Kingdom}
\date{\today}

\begin{abstract}
The possibility of using strongly and continuously interacting spins for quantum computation has recently
been discussed.
Here we present a simple optical scheme that achieves this goal while avoiding the drawbacks of earlier
proposals. We employ a third state,
accessed by a classical laser field, to create an effective barrier to information transfer.  The mechanism proves to be
highly efficient both for continuous and pulsed laser modes; moreover it is very robust, tolerating high decay
rates for the excited states. The approach is applicable to a broad range of systems, in particular dense structures such
as solid state self-assembled (e.g., molecular) devices. Importantly, there are existing structures upon which `first step'
experiments could be immediately performed.
\end{abstract}

\maketitle

The recent scientific literature contains an abundance of theoretical schemes for solid state quantum computation (QC).
Collectively
these schemes call for a quite bewildering range of physical systems.  
However, experimental progress has been slow because
it is hard to prevent decoherence {\em in an environment that supports controlled interactions} . Universal QC
requires both manipulation of individual qubits, and the ability to dictate their interactions with one another.
The former can (in many cases) be implemented by electromagnetic pulses that do not cause
severe decoherence. Control of interactions, however, is often supposed to be achieved by introducing some set of
control electrodes. Even in systems where this is physically possible, their presence directly adjacent to qubits may constitute a dangerous source of
decoherence. Moreover their use implies that the qubits are positioned at, or near, a complex patterned surface, inevitably associated with many decoherence channels.
While there do exist a variety of interesting schemes that implement interaction switching using an electromagnetic process, typically
these rely on either a conveniently passive interaction~\cite{NMRpaper} or else they require
directly taking the qubit out of its relatively safe subspace and into a space where it is subject to
decay~\cite{calarco03, nazir04}.

\begin{figure}[t]
\centering
\includegraphics[width=3.in]{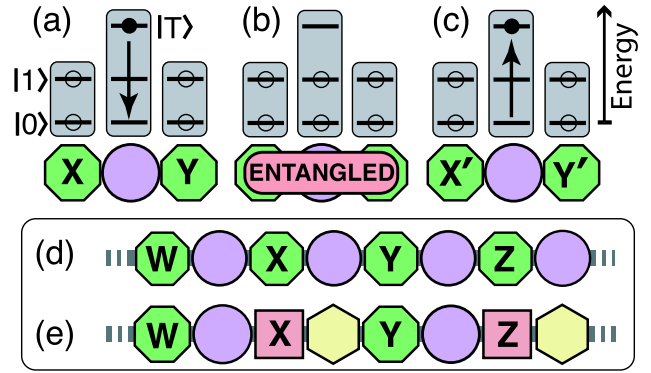}
\caption{(a)-(c) The simplest structure capable of demonstrating a qubit-qubit gate. There are three adjacent systems in a
linear
arrangement, with a permanent strong interaction existing between neighbors. Each system has two stable levels $\ket{0}$
and $\ket{1}$, and the central system has an additional third state $\ket{T}$. In (a) we suppose that the outer units
represent qubits $X$ and $Y$. The central system is acting as a barrier: in the simplest case it is `shelved' in state
$\ket{T}$, and is therefore far off-resonance from its neighbors. In practice the laser cycling mechanism depicted in
Fig.~\ref{figure2}
can form a superior barrier. In either case, when we wish to perform a gate operation we simply allow the
central system into its $\{\ket{0}$,$\ket{1}\}$ subspace. Under free evolution (b) it initially becomes entangled with
$X$ and $Y$, but `revives' to a product state at a certain time
later, allowing the barrier to be restored (c). The net effect is a unitary entangling gate on $X$ and $Y$.
Lower figure: (d) An $ABAB..$ pattern can suffice for a scalable architecture, but if global control~\cite{ababPRL} is
desired then a repeating set of four distinct systems (e) is appropriate.}
\label{figure1}
\end{figure}

All-optical manipulation of atomic systems has provided the basis for some of the most successful experiments
demonstrating aspects of QC~\cite{rauschenbeutel00, wineland04},
and here we shall adapt these ideas for use in solid state systems.
Our approach is a generalization of the barrier scheme of Benjamin and
Bose~\cite{benjamin04, longPaper}, which supported perpetual spin coupling by introducing a strong tuning of the
single-qubit Zeeman splitting. To perform such a tuning directly
would require
the equivalent of a qubit-by-qubit $\bf B$-field modulation, a serious experimental challenge.
Other comparable
schemes for perpetual coupling have equally significant drawbacks, such as the need for interaction switching during
initialization~\cite{Zhou}. Here we are able to dispense with these issues.

We shall begin by considering the smallest structure capable of realizing a two-qubit gate.
We envisage a set of three adjacent systems, in a linear arrangement, and in each we identify two states that are suitable
for qubit storage. We label those states $\ket{0}$ and $\ket{1}$. Within the Hilbert
space of these three two-state systems,
the Hamiltonian reads ${H}_{\rm comp}={H}_{\rm Z}+{H}_{\rm int}$, where ($\hbar = 1$)
\begin{eqnarray}
{H}_{\rm Z} & = & \sum_{j=1}^3 E_j{\sigma}^Z_j \\ \nonumber
{H}_{\rm int} & = & \sum_{j=1}^2 J_Z{\sigma}^Z_j{\sigma}^Z_{j+1}+
J_{XY}({\sigma}^X_j{\sigma}^X_{j+1}+{\sigma}^Y_j{\sigma}^Y_{j+1})   .
\end{eqnarray}
The $E_j$ are (time-independent) Zeeman
energies and
we have chosen a general (uniaxial) anisotropic interaction,
which has both the planar $XY$ interaction ($J_Z=0$) and the isotropic
Heisenberg interaction ($J_Z=J_{XY}$) as limits.
The simulations presented below will employ the $XY$ limit, but we have verified that the
same qualitative behavior is seen for any $J_Z$ of order $J_{XY}$. We now suppose that there is an additional level $\ket{T}$ available in the central
system~\cite{selective}. This state has an allowed transition to one of the system's qubit states; for clarity of
exposition we assume the transition is to state $\ket{0}$. We assume the transition energy $\hbar\omega_{0T}\gg kT$, where $T$ is the device's operating temperature. An on-resonance classical laser field will excite the transition via
${H}_{\rm laser}=  \Omega\cos(\omega_{0T} t) (\ket{T}\bra{0} + \ket{0}\bra{T})$, with $\Omega$ denoting
the Rabi frequency.

In the absence of the laser field, state $\ket{T}$ cannot become populated, and we effectively have
three coupled two-level
systems (Fig. 1) . Suppose that at some instant the left and right-hand systems each
represent a qubit, $X$ and $Y$, and the central system is in state $\ket{1}$. When this initial state evolves under
${H}_{\rm comp}$ we find that the three systems become mutually entangled. However, at a time
$t_R=\pi\hbar(8J_{XY}^2+J_Z^2)^{-\frac{1}{2}}$ the central system `revives'~\cite{longPaper} to state $\ket{1}$. At this
instant, the
transformation in the computational basis $\{\ket{00}$, $\ket{01}$, $\ket{10}$, $\ket{11}\}$ is given by (neglecting a global phase)
\begin{equation}
\label{2qgate}
U=\left(
\begin{array}{cccc}
1 &0 &0 &0\\
0 &iQs &Qc &0\\
0 &Qc &iQs &0\\
0 &0 &0 &W
\end{array}
\right).
\end{equation}
Here $Q=-\exp(i\phi)$, $s =\sin(\phi)$, $c = \cos(\phi)$ and $W=-\exp(-2i\phi)$, with
$\phi=\frac{\pi}{2}(1+8J_{XY}^2/J_Z^2)^{-\frac{1}{2}}$: $U$ is an entangling two-qubit gate, which
together
with arbitrary single qubit operations constitute a \emph{universal gate-set}~\cite{benjamin04,cirac01}.
The difficulty is that we must also be able to passivate the device by {\em effectively} decoupling the central barrier
before and after this free evolution. One might consider simply `shelving' the barrier into state $\ket{T}$ -
however this would require a high fidelity $\pi$ pulse
and, moreover, the third state would have to be stable (i.e., have a comparable lifetime to the $\ket{0}$, $\ket{1}$
states).
Typical real systems will not meet this condition.
However, as we shall now show, a rapid cycling of the central system to the excited state $\ket{T}$ can create an excellent barrier to qubit interaction,
even if $\ket{T}$ is unstable.
Thus the passive state\cite{phaseInPassive} of the device is to have the laser ON, and a gate is performed simply by pausing the
illumination for the period $t_R$.

In order to take account of the decay of $\ket{T}$, we
have employed two different approaches.
The first is a stochastic quantum jump method: the total system's
density matrix evolves coherently under ${H} = {H}_{\rm comp} + {H}_{\textrm{laser}}$ except that at random times the
population of
$\ket{T}$ collapses to $\ket{0}$.
The second approach is the Lindbladian formulation of the system's dynamics~\cite{mahler95}:
\begin{equation}
\label{mastereq}
\dot{\rho}=-i[H,\rho]+\frac{1}{2}\gamma
\left(2\sigma^-\rho\sigma^+-\sigma^+\sigma^-\rho-\rho\sigma^
+\sigma^-\right).
\end{equation}
$\sigma^+$ and $\sigma^-$ are the raising and lowering operators for the decaying transition, and $\gamma$ is
the decay rate. We
find that the two approaches are in excellent agreement~\cite{mapping}.

\begin{figure}[!b]
\centering
\includegraphics[width=3.4in]{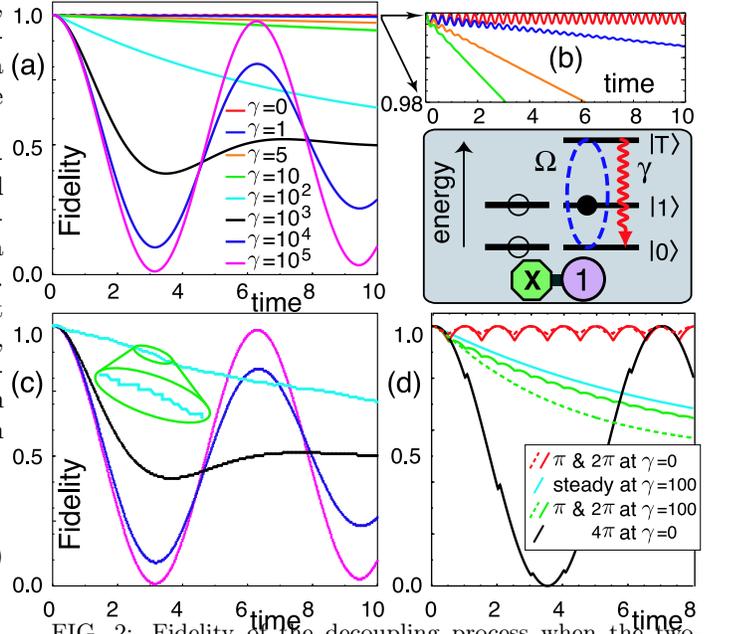}
\caption{Fidelity of the decoupling process when the two-spin system depicted in the schematic is subjected to laser
excitation. Time is in units of $1/J_{XY}$, and we have set $\omega_{01}=100J_{XY}$, $\omega_{0T}=1000J_{XY}$, and
the Rabi frequency $\Omega=40 J_{XY}$. (a) Simulation of continuous laser excitation using Eq.~\ref{mastereq}. Successive lines show the effect of
progressively greater decoherence rate $\gamma$ (in units of $J_{XY}$), and graph (b) `zooms' on the high fidelity region. For $\gamma$ values less than $\Omega$, i.e. decay weaker than the laser excitation rate, the decoupling is successful. In contrast, the
extremely high $\gamma$ plots expose the interesting phenomenon of {\em effective} laser deactivation, as described
in the text. In (c) we show the same numerical simulation, but now using the stochastic quantum jumps model for the decay. The agreement with (a) is seen to be very good.
Plot (d) shows the effect of a pulsed laser, with other parameters as in (a). We apply a repeating pattern of brief intense laser excitation followed by a
longer dark period, such that the average laser power is constant. We plot the effect when the pulse duration is chosen to effect
either a $\pi$, $2\pi$ or $4\pi$
rotation on the $\{\ket{0}$, $\ket{T}\}$ Bloch sphere.}
\label{figure2}
\end{figure}

Let us first consider a single qubit-bearing system coupled to a single barrier system, i.e., a
two-level system coupled to a single three-level
system (Fig. 2 schematic). This is the simplest system with which one could perform first `proof of principle' experiments.
In the numerical simulations depicted in Fig. 2, we first initialize the barrier system into state $\ket{1}$, and the
qubit-system in superposition
$\ket{\rm init}=\frac{1}{\sqrt2} (\ket{0}+\ket{1})$, so that the initial product state is
$\ket{1}\otimes\ket{\rm init}$. In the absence of a laser, the interaction $H_{\rm int}$ would cause
the $\ket{10}$ element of the superposition to exchange with $\ket{01}$, thus entangling barrier and qubit. To measure the
fidelity with which we can prevent this entanglement,  we calculate
${{\cal{F}}(t)}\equiv \bra{{\rm ideal}} \rho_Q(t)\ket{\rm ideal}$.
Here $\ket{\rm ideal}=\frac{1}{\sqrt2} (\ket{0}+\exp(- i \omega_{01}\tau)\ket{1})$, i.e., the
state for a perfectly isolated qubit, and $\rho_Q(t)$ is the qubit reduced density matrix.
Our choice of $\ket{\rm init}$ means that
if either the phase or amplitude of the `frozen' qubit deviate, then this is reflected in the fidelity measure. Adopting an equal superposition for $\ket{\rm init}$ is of course an arbitrary choice, however we have confirmed other initial
qubit states do not lead to significantly inferior fidelity.

In Fig. 2(a) the uppermost plot shows the simplest case of continuous laser, zero decay ($\gamma=0$). The fidelity
remains close to unity at all times.
This can be
explained by transforming the system to the dominant rotating frame, from which analysis one would
expect
${\cal{F}}=1-4(\Omega^2/J^2)\sin(\Omega t/2)$.
For
values of $\gamma$ within an order of magnitude of the laser Rabi frequency, we see
that fidelity follows the approximate trend ${\cal{F}} \simeq \frac{1}{2}[1+\exp(-t/k)]$
with $k\simeq 20\Omega/\gamma$. Given that in real systems it should be possible to excite the state
far faster than its spontaneous decay, this implies an excellent
degree of decoupling. 
However initial experiments may need to work with very non-ideal systems and so we also show the effect of large $\gamma$ values. For
$\gamma\gg\Omega$ the fidelity becomes oscillatory, and ultimately matches the case of zero decay, zero laser
power. This can be
understood by regarding the action of the decay as a quantum Zeno-effect~\cite{zenoEffect} which inhibits the laser transition
$\ket{0}\rightarrow\ket{T}$. If
a very rapid decay
mechanism could be switched, then one could elect to leave the laser on permanently and turn `on' the decay for a time $t_R$ to create a
perfect gate.
This observation serves to underline the
way that our scheme separates decoherence of $\ket{T}$ from decoherence of the qubit. Of course, the $X$ and $Y$ qubits will suffer the ``usual'' decoherence mechanisms that spins in solid state systems are subject to (except that we have done away with the need for gating electrodes, and indeed the need to be near a surface, and so certain mechanisms are removed). 
For our purpose the crucial criterion is that the native qubit lifetime be far longer than the characteristic timescale of our barrier mechanism, so that we can indeed neglect that lifetime in our analysis. Such systems exist, as we discuss later.

In Fig. 2(d) we depict the effect of a laser in pulsed mode.
Although we have demonstrated in the above analysis that we do not
{\em require} pulses to perform our decoupling, it is nevertheless interesting to investigate their effect, since a
first-step experimental study may well use a laser in this mode. We show the effect of a train of $\pi$,
$2\pi$ or
$4\pi$ pulses. Both the $\pi$ and $2\pi$ pulse trains maintain fidelity well, being slightly inferior to a continuous laser
with the same average power, while the $4\pi$ pulses have no effect. The mechanisms by which these pulses
cause decoupling are outlined in Fig. 3(a). Of course, a pulsed laser must necessarily contain a band of frequency components and so has a broader lineshape. In our simulations, the pulses we use are typically of duration $t_{pulse} = 0.1/J_{XY}$, which would imply a minimum laser bandwidth of order 10 $J_{XY}$. Any other energy levels which can be excited by the laser must be separated from the levels we are exploiting by at least this amount if the pulsed mode is to work. 

\begin{figure}[!b]
\centering
\includegraphics[width=3.in]{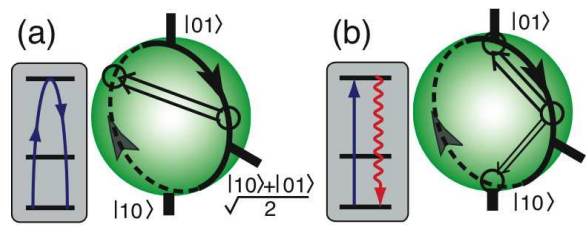}
\caption{An explanation of the effect of fast pulses seen in Fig. 2(d). We depict the Bloch sphere for the qubit-barrier
subspace $\{$$\ket{10}$,$\ket{01}$$\}$. The $XY$ interaction alone would simply
cause a rotation between these states - this is shown by the pole-to-pole
circle on the spheres. We suppose the initial state is $\ket{01}$, then after some small time $\delta t$ we apply a rapid
pulse to the barrier system. (a) A $2\pi$ pulse takes component $\ket{10}$ to $\ket{1T}$ and back, acquiring a net phase of
$-1$ so that we `short circuit' the Heisenberg cycle which then returns us $\ket{01}$. (b) The effect of a $\pi$ pulse plus
subsequent decay: provided $\delta t \ll 1/J_{XY}$ then the effect is a beneficial Zeno-type suppression~\cite{zenoEffect} of the
$\ket{01}\rightarrow\ket{10}$ transition. }
\label{figure3}
\end{figure}

Having examined the laser-driven decoupling process for the two-unit system (Fig.~\ref{figure2}), we confirmed
that precisely
the same decoupling effect occurs in the full three part qubit-barrier-qubit system~\cite{onlinegraphs}.

We now address the question of how our scheme could be realized in a specific system. A very promising possibility
is to embody the qubit in two Zeeman sublevels of a spin active nanostructure,
where $\ket{T}$ is a higher state
accessed by a laser of optical frequency. Pauli blocking~\cite{calarco03, nazir04} can lead to a spin-selective optical
excitation of excitons in quantum dots (QDs) with an excess spin.
Laser coupling to excitons
in arsenide QDs can be large enough to induce optical Rabi
oscillations
with a period of a few picoseconds~\cite{unold04, zrenner02},
which is at least three orders of magnitude longer than typical decoherence times in similar
structures~\cite{borri03, birkedal01}. This is more than sufficient for our scheme to work with extremely high fidelity.

Spin-spin interactions in semiconductor QDs can take several forms. In certain structures,
they might be due to the direct exchange mechanism, whose strength has been measured at 0.1~meV in
lithographically defined dots separated by about 200~nm~\cite{chen04}. Alternatively, they might be mediated by
conduction electrons (the RKKY interaction); suppression of Kondo resonances by this mechanism
has been observed recently~\cite{craig04}. These experiments suggest that gates could be performed on the sub-nanosecond
timescale, which is far shorter than measured spin relaxation times of more than 50~$\mu$s~\cite{hanson03}.
Though the lithographic dots offer a promising route to spin based QIP, we think that 
a first experiment testing our idea would use a pair of self assembled, Stranski-Krastanow dots, which have ideal optical properties.
The excess electron spins are provided by $n$-doping; layers of such dots have been fabricated, with an average of one electron per
dot, and spin polarization effects have been observed~\cite{cortez02} and spin lifetimes of 20~ms have been measured.~\cite{kroutvar04}
Polarized photoluminescence (PL) measurements may be used to verify the required electronic
structure, and by monitoring how the positions of the lines in PL spectra move when a tuned laser is
applied to the `blocking' transition, it should be possible to demonstrate the modulation of the spin-spin interaction.
State readout can be achieved by detecting the resonance fluorescence arising from the coupled nanostructure. In
particular, the readout of an electron spin in a QD can be accomplished by means of optical pumping with circularly
polarized light.~\cite{cortez02,efros03}.
This system could also be tested in pulsed laser mode, since the laser bandwidth needed to give the required pulse length should be about 1~meV. 
This energy is much less than the typical shell spacing of excitonic levels of small quantum dots, which is at least on the order of 10 meV~\cite{bayer00}. We do not need to be concerned with smaller spin splitting of levels, since our scheme would make use of selection rules governing the transitions made in circularly polarized light.

We could obtain a much stronger coupling between our qubits by using molecular structures. For example, the spacing between
spin-active endohedral fullerenes can be as small as 1~nm. Such structures also have extremely long spin decoherence
times of 50 $\mu$s~\cite{harneit02} and exhibit the magneto-optical activity which demonstrates the
presence of the required spin-dependent optical transition~\cite{morton}.

Single qubit operations can be performed very efficiently if the qubit units also possess a
higher state $\ket{T}$
(provided that their $0\leftrightarrow T$ splitting is well distinct from that of the barrier units)~\cite{calarco03}.
Instability of the higher states does not preclude their use in such operations, since a Raman
transition can be employed to avoid significantly populating the state~\cite{nazir04}.
Further, if a laser could be targeted to specific qubits, our device could be scaled by constructing a simple
repeating pattern like that of Fig. 1(d).
Alternatively, one
could employ the repeating $ABCD$ arrangement of Fig 1(e). Here there are two types of qubit system and two types of
barrier, where the distinction is in the $\omega_{0T}$ transition frequencies. Such a structure supports established
global control protocols~\cite{ababPRL}, making it suitable for molecular-scale structures where laser targeting of individual elements is impossible.

Ultimately, for a large scale device capable of outperforming classical machines, one would look for suitable ordered
two- or three-dimensional arrays~\cite{NJPpaper}. 
Perfectly ordered 2D metallic dot arrays, with alternating dot species, have been achieved~\cite{li02}.
These could either act as a template for further structures, or be directly charged to create atom-like states suited to
our protocol.
Engineered assemblies of CdSe nanocrystal QDs might also be used; these
have exhibited energy transfer on time scales of tens-of-picoseconds, while intrinsic
exciton lifetimes are $>20$~ns~\cite{crooker02}. In a basis where the presence or absence of an exciton is the pseudospin, this
interaction has the desired $XY$ form.
A similar interaction also exists in photosynthetic
bio-molecular complexes, which can exhibit energy transfer times of 0.3~ps~\cite{crooker02}.
These structures have recently been demonstrated to be
amenable to coherent quantum control~\cite{herek02} and could also constitute an
interesting implementation scenario.

In summary, we have shown that it is possible to perform scalable \emph{all-optical} QC using a solid state array of continuously and strongly interacting spins. The mechanism we employ is extremely robust and can operate with highly unstable states. Our detailed simulations have focused on the small scale structures, composed of two or three spins, that can be created with existing techniques. We have outlined `proof-of-principle' experiments that could be undertaken in the immediate future. 

We would like to thank J.~H.~Jefferson, J.~J.~L.~Morton, D. Jaksch and A. Beige for useful discussions. The work is supported by EPSRC (GR/R66029 and GR/S82176) and by the Royal Society.


\begin{references}
\bibitem{NMRpaper} J. A. Jones and E. Knill, J. Mag. Res. {\bf 141}, 322 (1999).
\bibitem{calarco03} T.~Calarco {\it et al.}, Phys. Rev. A {\bf 68}, 012310 (2003).
\bibitem{nazir04} A.~Nazir {\it et al.}, Phys. Rev. Lett. {\bf 93} 150502 (2004).
\bibitem{rauschenbeutel00} A.~Rauschenbeutal {\it et al.}, Science {\bf 288}, 2024 (2000).
\bibitem{wineland04} M.~D.~Barrett {\it et al.} Nature {\bf 429} 737 (2004).
\bibitem{benjamin04} S.~C.~Benjamin and S.~Bose,  Phys.~Rev.~Lett {\bf 90}, 247901 (2003).
\bibitem{longPaper} S. C. Benjamin \& S. Bose, preprint at

\noindent http://xxx.lanl.gov/abs/quant-ph/0401071.
\bibitem{ababPRL} S. C. Benjamin, Phys. Rev. Lett. {\bf 88}, 017904 (2002).
\bibitem{alluded} Ref.~\cite{benjamin04} did briefly allude to the possibility of using optical processes to cause an effective Zeeman shift - although the present schemes go well beyond that speculation.
\bibitem{Zhou} Zhou {\em at al}, Phys. Rev. Lett. {\bf 89}, 197903 (2002).
\bibitem{selective}It is necessary either that (a) there are two physically distinct systems, alternating, or (b) that the
laser can be selectively targeted to individual units along an identical chain. We assume the former.
\bibitem{phaseInPassive} Note that the dominant term in any additional interaction between $\ket{T}$ and the qubit states must be a simple Ising energy shift (energy transfer being negligible since $\omega_{0T}$ and $\omega_{01}$ are very far off resonance). It follows that pacification either by shelving to $\ket{T}$, or our rapid regular cycling, will generate only a known deterministic phase rotation which can be subsumed into the natural qubit phase accumulation. 
\bibitem{cirac01} K.~Hammerer  {\it et al.}, Phys.~Rev.~A {\bf 66}, 062321 (2002).
\bibitem{mahler95} G.~Mahler and V.~A.~Weberru{\ss}, {\it Quantum Networks} (Springer-Verlag, Berlin, 1995).
\bibitem{mapping} In Fig. 2 the decay rate $\gamma$ is mapped to an effective decay probability by comparing the graphs for one
set of parameters, then this same value is used in all other plots.
\bibitem{zenoEffect} P. Facchi {\em at al}, Phys. Rev. Lett. {\bf 86} 2699 (2001).


\bibitem{onlinegraphs}
The numerical experiment proceeded just as
before, except that now $\ket{\rm init}=\frac{1}{2}(\ket{00}+\ket{01}+\ket{10}+\ket{11})$, the two-qubit equal
superposition,
and similarly for $\ket{\rm ideal}$ and ${\cal{F}}$. Behavior was exactly as expected; please see 

\noindent http://www.nanotech.org/openResources/barrier.html
\bibitem{unold04} T.~Unold {\it et al.}, Phys.~Rev.~Lett. {\bf 92}, 157401 (2004).
\bibitem{zrenner02} A.~Zrenner {\it et al.}, Nature {\bf 418}, 612 (2002).
\bibitem{borri03} P.~Borri {\it et al.}, Phys.~Rev.~Lett.~{\bf 91}, 267401 (2003).
\bibitem{birkedal01} D.~Birkedal {\it et al.}, Phys.~Rev.~Lett {\bf 87}, 227401 (2001).
\bibitem{chen04} J.~C.~Chen {\it et al.}, Phys.~Rev.~Lett. {\bf 92}, 176801 (2004).
\bibitem{craig04} N.~J.~Craig {\it et al.}, Science {\bf 304}, 565 (2004).
\bibitem{hanson03} R.~Hanson {\it et al.}, Phys.~Rev.~Lett {\bf 91}, 196802 (2003).
\bibitem{bayer00} M.~Bayer, O.~Stern, P.~Hawrylak, S.~Fafard and A. Forchel, {\it Nature} {\bf 405} 923 (2004).
\bibitem{cortez02} S.~Cortez  {\it et al.}, Phys.~Rev.~Lett {\bf 89}, 207401 (2002).
\bibitem{kroutvar04} M.~Kroutvar {\it et al.} Nature {\bf 432} 012310 (2004).
\bibitem{efros03} A.~Shabaev   {\it et al.},  Phys.~Rev.~B {\bf 68}, 201305(R) (2003).
\bibitem{harneit02} W.~Harneit, Phys.~Rev.~A {\bf 65}, 032322 (2002).
\bibitem{morton} J.~J.~L.~Morton and M.~A.~G.~Jones, unpublished.
\bibitem{NJPpaper}S. C. Benjamin, New J. Phys. {\bf 6}, 61 (2004).
\bibitem{li02} J.~L.~Li {\it et al.}, Phys.~Rev.~Lett {\bf 88}, 066101 (2002).
\bibitem{crooker02} S.~A.~Crooker  {\it et al.}, Phys.~Rev.~Lett {\bf 89}, 186802 (2002).
\bibitem{herek02}J.~L.~Herek  {\it et al.}, Nature {\bf 417}, 533 (2002).
\end{references}
\end{document}